# Multitechnique characterization of Lapis Lazuli for provenance study

Alessandro Lo Giudice[1*], Alessandro Re[1], Silvia Calusi[2], Lorenzo Giuntini[2], Mirko Massi[2], Paolo Olivero[1], Giovanni Pratesi[3], Maria Albonico[1], Elisa Conz[1]

[1] *Dipartimento di Fisica Sperimentale, Università di Torino and INFN Sezione di Torino, Via Giuria 1, 10125, Torino, Italy*

[2] *Dipartimento di Fisica, Università and INFN Sezione di Firenze, Via Sansone 1, 50019, Sesto Fiorentino, Firenze, Italy*

[3] *Dipartimento di Scienze della Terra and Museo di Storia Naturale, Università di Firenze, Via G. La Pira 4, 50121 Firenze, Italy*

**ABSTRACT**

Lapis Lazuli is one of the oldest precious stone, being used for glyptic as early as 7000 years ago: jewels, amulets, seals and inlays are examples of objects produced using this material. Only few sources of Lapis Lazuli exist in the world due to the low probability of geological conditions in which it can form, so that the possibility to associate the raw material to man-made objects helps to reconstruct trade routes. Since art objects produced using Lapis Lazuli are valuable, only non-destructive investigations can be carried out to identify the provenance of the raw materials. Ionoluminescence (IL) is a good candidate for this task. Similarly to cathodoluminescence (CL), IL consists in the collection of luminescence spectra induced by MeV ion (usually protons) irradiation. The main advantage of IL consists in the possibility of working in air while measuring simultaneously the composition of major and trace element by means of complementary Ion Beam Analysis techniques like PIXE or PIGE (Particle Induce X-ray or Gamma-ray Emission).

In the present work a systematic study of the luminescence properties of Lapis Lazuli under charged particles irradiation is reported. In a first phase a multi-technique approach was adopted (CL, SEM with microanalysis, micro-Raman) to characterise luminescent minerals. This characterisation was propaedeutic for IL/PIXE/PIGE measurements carried out on significant areas selected on the basis of results obtained previously. Criteria to identify provenance of Lapis Lazuli from four of the main sources (Afghanistan, Pamir Mountains in Tajikistan, Chile and Siberia) were proposed.

*Keywords: Lapis Lazuli, Provenance, Ionoluminescence, Cathodoluminesce, Archaeometry, Ion Beam Analysis*

[*]Corresponding author: Tel. +390116707366, Fax. +390116691104, logiudice@to.infn.it

**INTRODUCTION**

Lapis Lazuli is one of the oldest precious stone, being used for glyptic as early as 7000 years ago [1]. It was diffused throughout the Ancient East and Egypt by the end of the fourth millennium B.C. and continued to be used during the following millennia: jewels, amulets, seals and inlays are examples of objects produced using this material. In more recent times, from sixth century B.C. until last centuries, finely grinded Lapis Lazuli was also used as pigment.

Only few sources of Lapis Lazuli exist in the world due to the low probability of geological conditions in which it can be formed [2] [3], so that the possibility to associate the raw material to man-made objects is helpful to reconstruct trade routes. This is especially true for ancient contexts where there is an absence or scarceness of written evidences [4]. Although the Badakhshan mines in Afghanistan (the more famous being Sar-e-Sang) are widely considered as now as the only sources of the Lapis Lazuli in ancient times [5] [6] [4] [7] other sources have been taken in consideration: Pamir mountains (Lyadzhuar Dara, Tajikistan) [5] [8] [9], Pakistan (Chagai Hills) [8] [9] [10], Siberia (Irkutsk, near Lake Baikal) [5], Iran [5] and Sinai [11]. The last two possibilities are not geologically confirmed and their interpretations are still debated [6] [9].

A systematic and exhaustive provenance study of the raw material utilized in works of art is still lacking. Results obtained using physico-chemical analysis (AAS) on limited quantities of cut wastes from Shahr-i-Sokta [8] [9] are in agreement with a Pamir mountains and Chagai Hills provenance, other than Badakhshan. Moreover, there is some evidence that also in Renaissance the Chagai Hills deposits could be utilized as source of Lapis Lazuli to produce ultramarine blue pigments [10]. Nevertheless Lapis Lazuli provenance is then a still open question and a systematic study on Lapis Lazuli objects would be helpful in resolving such problem. Pigments are less suitable for provenance study due to the processes involved in their fabrication that could introduce contaminants or remove minerals, even if good results were already reached [10] [12][13][14]. Since art objects produced using Lapis Lazuli are valuable, only non-destructive investigations can be carried out to identify the provenance of the raw materials. For this kind of investigation Prompt Gamma Activation Analysis (PGAA) was proposed and preliminary studies on rock samples were promising [15].

In this work, the capabilities of Ion Beam Analysis (IBA) to distinguish the provenance of Lapis Lazuli were investigated. IBA consists in the collection of different signals (visible photons, x-rays, gamma-rays and other) induced by MeV ion (usually protons) irradiation. Such analysis is not-destructive, it can be performed in air and sample pre-treatment is not required. Moreover by using a microbeam a spatial resolution of about 10 μm can be achieved, which is generally sufficient to distinguish mineral phases inside the samples. Particularly, this work is focused on Iono-Luminescence (IL), that appears as a good candidate for provenance studies. Similarly to the well-known cathodoluminescence (CL) extensively used in geosciences [16] [17] [18], IL consists in the collection of luminescence spectra induced by proton irradiation rather than electrons [19]. Besides general peculiarities of IBA techniques, such as the possibility to work in air, other advantages with respect to CL are: the greater probing depth (up to 0.1 mm) that allows the overcoming of surface layers on works of art; the possibility of performing simultaneous major and trace element analysis by means PIXE and PIGE. The IL capabilities to distinguish Chilean

Lapis Lazuli from Afghan and to apply this technique on works of art were already demonstrated in a previous preliminary work [20][21]. In this paper an in-depth study of the luminescence properties of Lapis Lazuli under charged particles irradiation is reported. Twelve stones coming from four of the main sources were systematically analysed using optical microscopy, cold-CL and SEM-EDS-CL techniques; in few cases micro-Raman spectroscopy was used to identify ambiguous phases. This characterisation work was propaedeutic to IL/PIXE/PIGE measurements. Such articulated use of complementary techniques is motivated by the less time required in measurements with respect to IBA techniques. In order to confirm the possibility of extending the results obtained by means CL to IL characterization, significative areas were selected and analysed by means of IL on the basis of results obtained using CL measurements.

**EXPERIMENTAL**

**Samples**

The Lapis Lazuli stones analysed in this work are part of the collection of the Mineralogy and Lithology section of the Museum of Natural History, University of Firenze [22]. In tab. 1 the catalogue number, a brief description of the starting material, the acquisition conditions and the provenances of the different samples are reported. As already pointed out by other authors [10][22], there is a lack of precise geographical information for many specimens. For most of the samples under investigation only the provenance area is specified, but the exact mines are not know: 3 samples come from Badakhshan in Afghanistan (probably Sar-e-Sang mine), 4 samples from America (only 2 samples are catalogued as Chilean, but probably are all from Ovalle), 1 sample from Irkutsk in Siberia, and 4 samples from Pamir mountains in Tajikistan (Lyadzhuar Dara lazurite deposit at 4800 m above see level and 76 km south of Khorugh in Gorno-Badakhshan).

**Instrumentation and operating conditions**

Samples were prepared as semi-thin sections (about 50 μm thick) and mounted on special slides with a 3 mm diameter hole in the centre. The holes were made to avoid any interference from the sample-holder during ion beam analysis, being the penetration depth of the proton probe of about 60-80 μm (SRIM calculation). It is worth stressing that this procedure for samples preparation was adopted only in the first characterization stage to simplify and speed up the measurements, but that it will be not necessary when works of art will be analysed by means of IBA techniques in future investigations.

All samples under study were observed in reflection mode by means of an optical microscope to evaluate their colour (Fig. 1) and to compare them with CL measurements. CL measurements were performed using two different instruments: a broad electron beam "cold cathode" apparatus and a SEM-CL system.

The cold cathode apparatus is a "CITL Cold Cathode Luminescence 8200 mk3" instrument equipped with a polarized optical microscope Olympus BH2 and a CCD camera. The semi-thin sections were placed in a vacuum chamber with transparent windows and the pressure was maintained at about 0.8 mbar. At this stage sample coating was not necessary because the working pressure was high enough to avoid charging effects. During the measurements, the accelerating voltage and current were 15 kV and 500 µA respectively. Since the electron beam was about 1 cm$^2$ broad, the current density was approximately 5 µA/mm$^2$. The penetration depth of the 15 keV electron beam probe is about 5 µm – 7 µm, depending on material density and composition. Lapis Lazuli showed an intense luminescence in the visible wavelengths so that acquisition time using standard camera conditions ranged from only 1 s to 20 s.

The SEM apparatus is a "Cambridge Stereoscan S360" equipped with Oxford PentaFET EDS and MONO-CL light collection system that can operate both in monochromatic (MC) or panchromatic (PC) mode. In MC mode the light from the sample is focused into a spectrometer consisting of a monochromator and a photomultiplier tube. The spectral range is about 300 – 800 nm and the wavelength resolution is of the order of few nanometers depending of monochromator slits aperture. The measured spectra were corrected for the spectral response of the instrument. In order to compare the relevant photons emissions, spectra were collected using the same operating conditions: working distance, magnification (5000×, scanning an area of about 20×25 µm$^2$), probe current (1 nA and 20 nA) and voltage (15 kV). CL maps were acquired mainly in MC mode owing to fluorescence phenomena. During CL-SEM analysis samples were not cooled at liquid nitrogen temperature to simulate measurement conditions on works of art. Due to the low working pressure slices were carbon coated to avoid surface charging.

Ion Beam Analysis was performed at external scanning micro-beam facility of the 3 MV Tandetron accelerator of the INFN LABEC Laboratory in Firenze [23]. Measurements were carried out using a 3 MeV proton beam (60-80 µm penetration depth), a current of about 200 pA and a lateral resolution of about 10 µm on 2×2 mm$^2$ areas. A motorized stage allows a wide range movement, therefore there are no limits in the analysed object dimension. Samples were characterized by means IL, PIXE and PIGE. IL apparatus was developed in the last two years mainly for cultural heritage applications. The spectral range is 200 nm – 900 nm and the resolution is 2 nm. The measured spectra were corrected for the spectral response of the instrument. More details on experimental set-up are reported in previous works [20] [21].

**RESULTS AND DISCUSSION**

Cold-CL images of the whole surface of all the samples were acquired using the above-mentioned experimental conditions; only camera integration time was modified during the measurements. In Fig. 2 is shown a representative collection of obtained images. Average luminescence intensity is similar in all samples except for the Siberian sample, which is more than five times less luminescent. From cold-CL images the distribution and intensity of luminescent mineral phases are clear. Minerals were identified mainly by means EDS analysis, but also micro-Raman was used. Moreover, PIXE analysis was performed in point of interest to evaluate trace elements.

With the exception of the Siberian sample, CL images are dominated by white-yellow luminescent phase that was identified as wollastonite in Chilean Lapis Lazuli and as diopside in the other samples. Another abundant common mineral phase shows a blue-azure luminescence; it was attributed to K-feldspars and it is the main feature of the Siberian sample. All the K-feldspars observed in lapis lazuli under investigation exhibit a lower luminescence with respect to diopside; this is the reason for the poor luminescence of this sample. Another significant feature is the orange-red colour that was associated to calcite crystals. Lazurite, which is responsible for the blue colour of Lapis Lazuli, shows a very weak luminescence and can be recognized as dark blue; for example in RI3065 the three regularly shaped dark regions are well formed lazurite crystals. The latter phase is dominant in Afghanistan samples. Lastly, pyrite do not exhibit any luminescence and therefore appears as black. Crystal dimensions vary from few micrometers to hundred of micrometers but do not seem to be related to different provenances. Apart from wollastonite founded only in Chilean samples, all the above-mentioned phases were found in different proportions in all the samples.

In order to acquire data with significant statistics, ~~many~~ spectra for each luminescent phase were collected in all the samples by means of CL-SEM, for a total of more than a hundred acquisitions. Many spectra exhibit a strong dependence on beam current density and therefore such parameter was kept at about 2-40 $\mu A/mm^2$, similarly to the value adopted in cold-CL analysis.

Spectra from calcite and K-feldspar (not reported here) exhibit the same features in all the samples. Calcite spectra are characterized by two broad bands at 625 nm and 350 nm. These bands are due to $Mn^{2+}$ substituting $Ca^{2+}$ and to intrinsic luminescence, respectively [24]. In K-feldspar spectra the only band observed was at 455 nm; such emission is very common in this mineral due to $Al-O^--Al$ defects [25].

In all the samples, lazurite has a very weak luminescence and spectra show the same bands of the main luminescent phases, i.e. wollastonite in Chilean, K-feldspar in Siberian and diopside in the other samples. This can be attributed to the contaminations of these phases inside the lazurite or to luminescence generated by multiple scattering during measurements (i.e. electrons hit strong luminescent phases after backscattering from the detectors). In any case, such luminescence is too weak to be useful for any provenance study.

Wollastonite ($CaSiO_3$) is one of the strongest luminescent phases and it is common to all Chilean samples. Spectra are characterized by three bands (Fig. 2) at about 450 nm, 560 nm and 620 nm. The first band is characteristic of silicate matrix and is common to other phases such as diopside [26]. The other two closed bands (more evident in Fig. 5, where the corresponding IL spectra are shown) are related to α-$CaSiO_3$:Mn and β-$CaSiO_3$:Mn polymorphic structures [27]. Such manganese-related bands were not observed in any other samples and therefore represent a significant fingerprint for Chilean provenance, as already observed in [20] [21].

With the exception of Chilean samples, diopside ($CaMgSi_2O_6$) was found in all Lapis Lazuli and together with wollastonite, it is the strongest luminescent phase. In Fig. 3 diopside spectra from few samples are shown in comparison with wollastonite spectrum of a Chilean sample. Apart from the 450 nm and the infrared emission (probably due to Fe-related impurities [28]), the spectra are characterized by an intense band at about 580 nm and by a weaker band at 690 nm. These two bands can be attributed to $Mn^{2+}$ ions in M1 sites and in M2 sites [29]. The 690 nm emission was observed only in Pamir samples and therefore it could represent a good criterion to distinguish this provenance from the others.

Lastly the UV detection capabilities of the CL-SEM apparatus allowed the observation of a luminescent phase that was not visible in cold-CL images due to the low camera sensitivity in this wavelength range. From EDS/PIXE composition and preliminary micro-XRD analysis, this mineral phase is probably related to the cancrinite group. A CL-SEM image obtained at 350 nm (10 nm pass band) is shown in Fig. 4 in comparison with a cold-CL image of a 200 μm cancrinite crystal. The CL spectrum (also shown in Fig.4) is very interesting and can be attributed to the strong UV emission and to a vibronic structure with ZPL at 2.55 eV. This phase was abundantly observed only in Pamir samples and therefore it could be used to distinguish this provenance from the others.

In order to confirm the possibility of extending the CL results to IL, a number of crystals were selected to be analysed by means of IBA techniques. Spectra obtained using CL and IL were compared, exhibiting a good correlation. As an example, Fig. 5 reports the spectra collected from wollastonite in a American samples, and from diopside and cancrinite in a Pamir sample.

A systematic PIXE study was not performed even if trace elements inside phases could be useful for provenance. Arsenic in Pamir lazurite and zirconium in Afghanistan lazurite could be good candidates for thus purpose. At present, significant PIXE information come from barium and strontium contents. In fact all the lapis lazuli samples show a barium contents that is below the instrument sensitivity, with the exception of the Siberian sample, in which the average content is higher than 1%. Also strontium concentrastion is more than ten times higher in Siberian lapis lazuli (6000 ppm) in comparison with other provenances. These results confirm previous works on Siberian lapis lazuli [9][30] and can be used to distinguish this provenance from the others.

**CONCLUSIONS**

In this work we reported a systematic study on the luminescent properties of Lapis Lazuli under charged particle irradiation from four of the main sources. Despite the limited number of analyzed samples, preliminary results are promising and experimental differences among different samples are meaningful. Spectra obtained by means CL and IL were compared showing a good correlation so that the possibility of extending the results obtained by means CL to IL was confirmed.

Differences in luminescence spectra arise from the presence of peculiar mineral phases associated with Lapis Lazuli sources and could be used to distinguish the provenance of the stones.

The main distinctive mineral phase in Chilean Lapis Lazuli is wollastonite, which can be distinguished by the intense luminescence at 560 nm and 620 nm. This behaviour was observed in 4 samples and confirms what was obtained in previous works [20] [21].

Lapis Lazuli from Pamir exhibits an additional luminescence band at 690 nm, corresponding to the emission of diopside. It is also characterized by a cancrinite phase with a strong UV emission and a vibronic structure with ZPL at 2.55 eV.

Although differences exist in cold-CL images and relevant CL spectra from Afghanistan and Siberian samples, they are weak because they are based only on different mineral phases ratio. However, Siberian Lapis Lazuli is characterized by a high contents of Barium as already shown by other authors [9][30] and therefore the simultaneous PIXE analysis allows the discrimination of this provenance.

It is worth stressing that all results in this work are related to a limited quantity of samples, and that the study will be tested with increasing statistics in future works. If the above mentioned criteria will not be confirmed, new features will be investigated, such as for example the dependence of the spectra from the current density.

**ACKNOWLEDGEMENTS**

This work has been carried out and financially supported in the framework of the INFN experiments ''DANTE" and "FARE". The authors gratefully acknowledge the "Compagnia di San Paolo" for the funding of a PhD scholarship on Physics applied to Cultural Heritage. The authors warmly thank Prof. Luca Martire (Department of Earth Science, University of Torino) for the availability of his cold-CL apparatus.